
\input phyzzx
\def\ri{\rightarrow}
\def\cl{{\cal L}_W}
\def\clt{{\cal L}^{(2)}}
\def\clf{{\cal L}^{(4)}}
\def\tr{{\rm Tr}}
\def\ri{\rightarrow}
\def\clnon{{\cal L}^{\Delta S=1}_{\rm non-anom}}
\def\clano{{\cal L}^{\Delta S=1}_{\rm anom}}
\def\kpiee{K^+\ri\pi^+ e^+ e^-}
\def\nc{$N_c~$}
\def\br{{\rm Br}}
\def\de{{_{\rm DE}}}
\def\GeV{{\rm GeV}}

\def\sq2{\sqrt{2}}

\def\di{\Delta I={1\over 2}}
\def\dit{\Delta I={3\over 2}}

\def\cl{{\cal L}_W}

\def\sq2{\sqrt{2}}

\pubnum={17-92}
\date{October, 1992}
\pubtype{}
\titlepage
\title
{Large-$N_c$ Higher Order Weak Chiral Lagrangians
for Nonleptonic and Radiative Kaon Decays$^*$}
\hoffset=0pt
\hsize=6.5in
\voffset=0in
\vsize=9.0in
\author
{Hai-Yang Cheng$^\dagger$}
\address{Institute of Physics, Academia Sinica, Taipei, Taiwan
11529, Republic of China}
\abstract
    In pure chiral perturbation theory (ChPT) the couplings of higher
order Lagrangian terms are running parameters and hence can be
determined only empirically from various low-energy hadronic processes.
While this scenario works well for strong interactions, it is
unsatisfactory for nonleptonic and radiative nonleptonic weak interactions:
It is impossible, from the outset, to determine all unknown
higher-order chiral couplings by experiment; theory is tested only by
certain chiral constraints rather than by its quantitative predictions.
Based on a QCD-motivated model for $p^4$ strong chiral
Lagrangian valid in the limit of large $N_c$,
one can derive large-$N_c$ fourth-order effective
chiral Lagrangians for $\Delta S=1$ nonleptonic weak interactions and
radiative weak transitions. Applications to
$K\ri\pi\pi\pi,~K\ri\pi\gamma\gamma$, and $K\ri\pi\pi\gamma$
are discussed.
\vskip 2.0 cm
\noindent $^*$ Invited talk presented at The Second German-Chinese
Symposium on Medium Energy Physics, Sept. 7-11, 1992, Bochum, Germany.

\noindent $^\dagger$ Bitnet address: phcheng@twnas886, hcheng@bnlcl1.
\endpage

\noindent {\bf 1. Introduction}

   There are at least two reasons that lead us to study seriously the
structure of higher-order effective chiral Lagrangians for weak
interactions. First, the $K(k)\ri\pi(p_1)\pi(p_2)\pi(p_3)$ decay amplitude
in the Dalitz plot is conventionally parametrized as
$$A(K\ri 3\pi)=\,a+bY+c(Y^2+X^2/3)+d(Y^2-X^2/3),\eqno(1.1)$$
where $Y=(s_3-s_0)/m^2_\pi,~X=(s_2-s_1)/m^2_\pi,~s_i=(k-p_i)^2,~s_0=
(s_1+s_2+s_3)/3.$ The experimental signal for the quadratic terms ($i.e.,$
the parameters $c$ and $d$) requires the inclusion of higher order
weak Lagrangian terms containing four or more derivatives. Also, it is
well known that current-algebra predictions for $a$ and $b$ are too
small by $18\%$ and $35\%$ respectively. Second, the radiative kaon
transition cannot be generated by the lowest-order chiral Lagrangian
since Lorentz and gauge invariance requires at least two powers of
momenta in the radiative decay amplitude. Therefore, it is necessary of
higher order in chiral perturbation theory.

   The couplings of higher-order chiral Lagrangians depend on the choice
of the renormalization scale $\mu$ as divergences of chiral loops are
absorbed by the counterterms which have the same structure as higher
derivative chiral terms. Consequently, the unknown running parameters are
not really fundamental coupling constants; also they
cannot be fixed by the requirement of chiral symmetry alone.
For strong interactions, Gasser and Leutwyler\refmark{1} have empirically
determined the coupling parameters at the mass scale $\mu=m_\eta$ from
various low-energy hadronic processes in conjunction with the
Zweig-rule argument.

   Despite the fact that pure ChPT is phenomenologically successful
when applied to low-energy strong-interaction physical processes,
this approach is unsatisfactory for nonleptonic and radiative
weak interactions. For example, in the chiral limit, there are
seven independent $p^4$ Lagrangian $\clf_W$ responsible for $\Delta
S=1$, $\Delta I={1\over 2}$ weak transitions. Unfortunately, there is
only one process, namely $K\ri 3\pi$ decay, relevant for the
determination of $\clf_W$. Unlike the strong interaction,
it is impossible from the outset to completely determine the structure
of $\clf_W$; only certain chiral constraint relations can be tested
to check the validity of ChPT at the four-derivative level.\refmark{2}
That is, although pure ChPT is a rigorous theory, its ability of making
predictions for phenomenological $p^4$ weak transitions and for
radiative nonleptonic decays is rather limited.
While the unknown coupling constants in ChPT in principle must be
determined by experiment so that its prediction is truly model
independent, it is also important to appeal a dynamic model to help
us understand the underlying physics if theory by itself does not lead to
any siginificant quantitative predictions owing to present limitation
from both experiment and theory.

    A QCD-inspired model for nonanomalous four-derivative effective
action for strong interactions does exist in the large $N_c$
limit.\refmark{3-6}
It is obtained by coupling QCD to external meson and gauge fields.
In the limit of $N_c$, chiral loops are suppressed and the effects
of interest are due to the quark loops and the gluonic corrections
arising from all planar diagrams without the internal quark loop.
In the leading $1/N_c$ approach, coupling constants become
renormalization scale independent and are (almost) uniquely determined.

One can then proceed to derive a large \nc effective Lagrangian for
nonleptonic $\Delta S=1$ weak interactions at order $p^4$ based on the
following three ingredients\refmark{7}: a rather
simple structure of the effective weak Hamiltonian in the leading $1/N_c$
expansion, bosonization up to the subleading order, and factorization
valid in the limit of large $N_c$. Confrontation with experiment
for $K\ri 3\pi$ decays reveals a good agreement for two of
the measured parameters in the Dalitz expansion of $K\ri 3\pi$
amplitudes.\refmark{8} Based on the same approach, a derivation of
non-anomalous and anomalous fourth-order chiral Lagrangians
responsible for $\Delta S=1$ radiative weak transitions is
also straightforward.\refmark{9}
\vskip 0.7 cm
\noindent {\bf 2. Higher-order Chiral Lagrangians for Nonleptonic Weak
Interactions}

   The lowest order chiral Lagrangian including explicit
chiral-symmetry breaking for low-energy QCD is given by\refmark{10}
$$\clt_S={f_\pi^2\over 8}\tr(\partial_\mu U\partial^\mu U^\dagger)
+{f^2_\pi\over 8}\tr(MU^\dagger+UM^\dagger),\eqno(2.1)$$
where
$U=\exp(2i{\phi\over f_\pi}),~~\phi={1\over\sqrt{2}}\phi^a\lambda^a,~~
\tr(\lambda^a\lambda^b)=2\delta^{ab},~~f_\pi={\rm 132~MeV}$,
and $M$ is a meson mass matrix with the non-vanishing matrix elements
$M_{11}=M_{22}=m^2_\pi,~~M_{33}=2m^2_K-m^2_\pi$.
It was established by Gasser and Leutwyler\refmark{1} that in the chiral
limit the most general expressions for the $p^4$
effective chiral Lagrangians including external vector $V_\mu$ and
axial-vector $A_\mu$ gauge fields are
$$\eqalign{\clf_S= & L_1[\tr(D^\mu U^\dagger D_\mu U)]^2+
L_2[\tr(D_\mu U^\dagger D_\nu U)]^2+L_3\tr(D^\mu U^\dagger D_\mu U)^2
\cr  - &\,iL_9\tr(F^R_{\mu\nu}D^\mu U^\dagger D^\nu U+F^L_{\mu\nu}
D^\mu UD^\nu U^\dagger)+L_{10}\tr(U^\dagger F^R_{\mu\nu}UF^{\mu\nu L}
) \cr}\eqno(2.2)$$
with
$$\eqalign{ D_\mu U= &\,\partial_\mu U+A^L_\mu U-UA^R_\mu, \cr
F^{L,R}_{\mu\nu}= &\,\partial_\mu A^{L,R}_\nu-\partial_\nu A^{L,R}_\mu
+[A^{L,R}_\mu,~A^{L,R}_\nu], \cr
A^{L,R}_\mu= &\,V_\mu\pm A_\mu.  \cr}\eqno(2.3)$$
Gasser and Leutwyler have
empirically determined the parameters $L_1,...L_{10}$
at the mass scale $\mu=m_\eta$
from various low energy hadronic processes in conjunction with the Zweig-rule
argument.

In the limit of large $N_c$ (\nc being the number of colors), the
aforementioned chiral couplings are theoretically manageable at least
to the zeroth order in $\alpha_s^2$. There exist several approaches
for the computation of $L_i$.\refmark{3-6} Here, we will only mention a
formal one.\refmark{3,5}
First of all, the chiral-loop contribution is suppressed by at least
a factor of $1/N_c$ relative to the quark loop at the same order
of $p^n$ in the leading $1/N_c$ expansion. Subsequently, the
higher order couplings in the large \nc chiral perturbation theory
 are renormalization
scale independent. Second, consider QCD coupled to external gauge
fields. The integration of both quark and gluonic degrees of freedom
yields two different categories of global chiral anomalies: proper
(Bardeen) anomalies which contain the totally antisymmetric tensor
$\epsilon_{\mu\nu\alpha\beta}$ and spurious anomalies which do not.
Now the variation of the generating function under a local chiral
transformation is governed by chiral anomalies. It is well known that
the integration of topological anomalies gives the Wess-Zumino-Witten
effective action. Likewise, the integration of nontopological
anomalies yields an action for $p^4$ nonanomalous chiral Lagrangians.
A consistent leading $1/N_c$ expansion requires one to include
not only the contributions from the quark loops but also the gluonic
effects arising from all planar diagrams without the internal quark
loops. The gluonic corrections which have been neglected in previous
publications were dicussed in ref.[5]. In the chiral limit, the
large-$N_c$ chiral couplings valid to the
leading order in gluonic corrections read\refmark{3-6}
$$8L_1=4L_2=L_9={N_c\over 48\pi^2},~~~~L_3=L_{10}={N_c\over 96\pi^2}(
1+\xi).\eqno(2.4)$$
It was shown in ref.[5] that to the first order in $\alpha_s$ only
the couplings $L_3$ and $L_{10}$ receive gluonic modeifications
denoted by the parameter $\xi$ in (2.4).
\foot{There exists an inconsistency for the chiral coupling $L_{10}^r$.
The value
of $L_{10}^r(\mu=m_\rho)=-(5.2\pm 0.3)\times 10^{-3}$ often quoted in the
literature is obtained from the experimental measurement of the
axial-to-vector form factor ratio $f_A/f_V=0.52\pm 0.06$ (The
updated value is $0.45\pm 0.06$\refmark{11}) in the radiative pion decay.
On the other hand, it is extracted to be $-(2.9\pm 0.8)\times 10^{-3}$
from the pion polarizability measured in the pion Compton
scattering.\refmark{12} While the former value is based on more precisive
experiments, the latter is favored by theory (though some experts may
disagree on this point) for the reason that both $L_3$ and $L_{10}$
receive the same amount of gluonic corrections\refmark{5} and that the
predicted $L_3=3.2\times 10^{-3}$ (note that $L_3$ is $\mu$-independent)
to the zeroth order in $\alpha_s$ is already in
good agreement with experiment. It is thus not clear to us what is
the origin for the discrepancy. Fortunately, the coupling $L_{10}$
is not relevant for our ensuing discussion. Therefore, we can safely
put $\xi=0$ for our later purposes.}

    Several remarks are in order. (i) The strong effective Lagrangian
given by Eq.(2.4) should be viewed as a QCD-motivated model rather than
a formal chiral Lagrangian derived from large-$N_c$ QCD: It is obtained
by coupling QCD to external gauge fields and considering its anomalous
variation. (ii) Since the chiral parameters $L_i$ in the leading
$1/N_c$ expansion are scale-independent constants, one should in
principle {\it not} compare Eq.(2.4) with the running couplings
$L_i(\mu)$ determined from experiment. Nevertheless, empirically they are
quite similar at the mass scale between 0.5 and 1 GeV.\refmark{13}

It is known that the lowest-order chiral Lagrangian responsible
for $\Delta S=1$ and $\Delta I={1\over 2}$ nonleptonic weak
interactions reads
$$\clt_W=-g_{_8}\tr(\lambda_6 L_\mu L^\mu),\eqno(2.5)$$
where $L_\mu\equiv(\partial_\mu U)U^\dagger$ is an $SU(3)_R$ singlet
and $L^\dagger_\mu=-L_\mu$. The parameter $g_{_8}$ of the octet
weak interaction is determined from the measured $K\ri\pi\pi$ rates.
In the chiral limit and in the absence of external gauge fields there
are seven independent CP-even quartic-derivative weak Lagrangian
terms\refmark{14} which transform as ($8_L,\,1_R$) under chiral
rotations:
\foot{For the construction of the most general expressions of the
counterterm Lagrangians relevant for the nonleptonic weak interactions,
see ref.[15].}
$$\eqalign{ \clf_W= &\,{g_8\over f^2_\pi}\bigg\{ h_1
\tr(\lambda_6 L_\mu L^\mu
L_\nu L^\nu)+h_2\tr(\lambda_6 L_\mu L_\nu L^\mu L^\nu)  \cr
+ &\,h_3\tr(\lambda_6 L_\mu L_\nu L^\nu L^\mu)+h_4\tr(\lambda_6
L_\mu L_\nu)\tr(L^\mu L^\nu)  \cr   + &\,h_5\tr(\lambda_6\tilde{Y}
\tilde{Y})+h_6\tr\left([\lambda_6,~\tilde{Y}]L_\mu L^\mu\right)
+h_7\tr\left([\lambda_6,~\tilde{Y}_{\mu\nu}]L^\mu L^\nu\right)
\bigg\},  \cr}\eqno(2.6)$$
where $Y_{\mu\nu}=(\partial_\mu\partial_\nu U)U^\dagger,~\tilde{Y}
_{\mu\nu}=Y_{\mu\nu}-Y_{\mu\nu}^\dagger$, and $\tilde{Y}=g^{\mu\nu}
\tilde{Y}_{\mu\nu}$. Under the CP transformation, $\tilde{Y}_{\mu\nu}
\ri-\tilde{Y}_{\mu\nu}^T$.

To determine the weak chiral parameters $h_i$ in the $1/N_c$ expansion
requires three ingredients: the $\Delta S=1$ effective weak Hamiltonian
at the quark level, bosonization and factorization, as we are going
to elaborate on. The $\Delta S=1$ effective nonleptonic Hamiltonian
in the limit of large \nc has a rather simple structure\refmark{16}
$$\eqalign{ {\cal H}^{\Delta S=1}_{\rm eff}= &\,{G_F\over\sqrt{2}}
\sin\theta_{_C}\cos\theta_{_C}\left\{c_8(Q_2-Q_1)+c_{27}(Q_2+2Q_1)
\right\}, \cr Q_1= &\,(\bar{s}d)(\bar{u}u),~~~Q_2=\,(\bar{s}u)(\bar{u}
d), \cr}\eqno(2.7)$$
where $(\bar{q}_iq_j)\equiv\bar{q}_i\gamma_\mu(1-\gamma_5)q_j$. The
combination $(Q_2-Q_1)$ is a $\Delta I={1\over 2}$ four-quark operator
which transforms as ($8_L,\,1_R$) under chiral rotation, while $(Q_2+
2Q_1)$ is equivalent to a 27-plet $\Delta I={3\over 2}$ operator
in the large-$N_c$ approximation.

    Using Eqs.(2.2) and (2.4) one can determine the bosonization of the
quark current $J_\mu^{ij}\equiv (\bar{q}^iq^j)$ to the next-to-leading
order in chiral expansion. Writing $(J_\mu)_{ji}=(if_\pi^2/2)(\hat{L}_\mu
)_{ij}$, the result is
$$\hat{L}_\mu=L_\mu+{N_c\over 24\pi^2 f_\pi^2}\left\{\,(L_\nu L^\nu L_\mu+
L_\mu L_\nu L^\nu)+[\tilde{Y},~L_\mu]+[L^\nu,~\tilde{Y}_{\nu\mu}]\right\}.
\eqno(2.8)$$
Since factorization is valid in the large-$N_c$ approximation, one may
substitute (2.8) into (2.7) to obtain the quartic-derivative weak chiral
couplings $h_i$\refmark{7}
$$h_1=-{1\over 3}h_2={1\over 3}h_4=h_6=-h_7=\,{N_c\over 24\pi^2},~~~~
h_3=h_5=0.\eqno(2.9)$$
The effective weak chiral Lagrangians $\clt_W+\clf_W(1/N_c)$ have been
tested successfully in the study of the nonleptonic $K\ri\pi\pi\pi$
decay.\refmark{8}

\vskip 0.7 cm
\noindent {\bf 3. Electromagnetically Induced Anomalous and Non-anomalous
Weak Chiral Lagrangians}

The most general $p^4$ electromagnetically induced $\Delta S=1$
non-anomalous weak Lagrangians with at most two external photon fields
which satisfy the constraints of chiral and CPS
symmetry have the form
$$\eqalign{\clnon= &\,i\left({2\over f^2_\pi}\right)g_8eF^{\mu\nu}
\left[\,\omega_1\tr(\lambda_6 L_\mu L_\nu Q)+\omega_2\tr(\lambda_6
L_\nu QL_\mu)\right]  \cr + &\,\omega_3\left({2\over f^2_\pi}\right)g_8
e^2F^{\mu\nu}F_{\mu\nu}\,\tr(\lambda_6 QUQU^\dagger),
\cr}\eqno(3.1)$$
while the  anomalous terms are $(\tilde{F}_{\mu\nu}\equiv\epsilon_{\mu\nu
\alpha\beta}F^{\alpha\beta})$
\foot{Note that our $\omega_3,~\omega_4$ are the couplings $\omega_4,~
\omega_3$ respectively in ref.[17]. The last three terms in (3.2) are
missed in the same reference.}
$$\eqalign{ \clano= & i\omega_4\left({2\over f^2_\pi}\right)g_8e\tilde{F}
^{\mu\nu}\tr(QL_\mu)\tr(\lambda_6 L_\nu)  \cr
+& i\omega_5\left({2\over f^2_\pi}\right)g_8e\tilde{F}^{\mu\nu}\tr(QU^\dagger
L_\mu U)\tr(\lambda_6L_\nu)  \cr
+& i\omega_6\left({2\over f^2_\pi}\right)g_8e\tilde{F}^{\mu\nu}\tr\left(
\lambda_6[UQU^\dagger,~L_\mu L_\nu]\right)  \cr
+& \omega_7\left({2\over f^2_\pi}\right)g_8e^2F^{\mu\nu}F_{\mu\nu}
\epsilon^{\alpha\beta\rho\sigma}\tr(\lambda_6L_\alpha)\tr(L_\beta L_\rho
L_\sigma),   \cr}\eqno(3.2)$$
in which the ordinary derivative in $L_\mu$ is replaced by the covariant
derivative in the presence of external gauge fields.
Presently, there are only two information on the unknown parameters
$\omega_i$. First of all, Ecker, Pich and de Rafael (EPR)\refmark{17} found
the relation $\omega_2=\,4L_9$, which
must hold at least for the divergent parts of the counterterm
coupling constants because they must render the divergent loop
amplitudes finite. Second, from the BNL measurement of the
$\kpiee$ decay rate,\refmark{18} one finds a scale-independent relation
$$\omega_1+2\omega_2-12L_9\simeq \,-7.5\times 10^{-3}.\eqno(3.3)$$
This together with the empirical value of $L_9^r(\mu=m_\rho)=6.7\times
10^{-3}$ leads to
$$\omega_1^r(\mu=m_\rho)+2\omega_2^r(\mu=m_\rho)\simeq 0.074~.\eqno(3.4)$$

    In the presence of the external electromagnetic field, the gauge
field in Eq.(2.3) is identified with
$$A_\mu^L=-ieA_\mu Q,~~~~A_\mu^R=-ieA_\mu Q,\eqno(3.5)$$
with $Q={\rm diag}(2/3,-1/3,-1/3)$ and $A_\mu$ being the photon field.
Just as in Sec.II, one first finds out
the bosonization of the quark current in the presence of external
photon field and then substitutes it into the four-quark operator
$(Q_2-Q_1)$ and gets the non-anomalous
$\Delta S=1$ Lagrangian $\clnon$ [Eq.(3.1)] with
\foot{Recently, very different large-$N_c$ predictions $\omega_1=\omega_2
=8L_9,~\omega_3=12L_{10}$ were obtained by Bruno and Prades.\refmark{19}
This is attributed to the fact that the effect of the quark operator $Q_1$
is not considered by them.}
$$\omega_1=\omega_2=\,{N_c\over 12\pi^2},~~\omega_3=0.\eqno(3.6)$$
The previous observation of $\omega_2=4L_9$ made by EPR is
reproduced here. Evidently, the large-$N_c$ prediction of $\omega_1+
2\omega_2=0.076$ is remarkably in agreement with (3.4).
\foot{The large-$N_c$ prediction $\omega_1+2\omega_2-12L_9=0$ is also
in good agreement with (3.3) in view of the fact that $12L^r_9(\mu=m_\rho)
\simeq 0.08$.}

   The derivation of the large-$N_c$ anomalous weak chiral Lagrangian
coupled to external photon fields is more complicated but more
interesting as it is governed by chiral anomalies.
To do this, one first writes down
the relevant Wess-Zumino-Witten terms
$$\eqalign{ {\cal L}_{_{\rm WZW}}=\,- & {N_c\over 48\pi^2}\epsilon
^{\mu\nu\rho\sigma}\tr\big\{-(A^R_\mu R_\nu R_\rho R_\sigma+A^L_\mu
L_\nu L_\rho L_\sigma)  \cr  - &{1\over 2}A^L_\mu L_\nu A^L_\rho
L_\sigma-A^R_\mu U^\dagger A^L_\nu UR_\rho R_\sigma +A^L_\mu U
A^R_\nu U^\dagger L_\rho L_\sigma  \cr + &\partial_\mu A^R_\nu
U^\dagger A^L_\rho UR_\sigma+\partial_\mu A^L_\nu UA^R_\rho
U^\dagger L_\sigma  \cr + &(A^L_\mu\partial_\nu A^L_\rho+\partial
_\mu A^L_\nu A^L_\rho)L_\sigma\big\}+...,  \cr}\eqno(3.7)$$
where $R_\mu\equiv U^\dagger\partial_\mu U$.
Once the bosonization in the anomalous case is found after a lengthy
manipulation, it is straightforward to show that
\foot{It was wrongly conjectured in ref.[20] that the couplings
$\omega_4,...,\omega_7$ have nothing to do with the chiral anomaly.
This has been corrected in ref.[21] and is now consistent with ref.[9].}
$$\omega_4=2\omega_5=4\omega_6=-8\omega_7={N_c\over 12\pi^2}.\eqno(3.8)$$
This result first obtained in ref.[9] was recently confirmed by
ref.[21]. It should be stressed that the anomalous chiral coupling constants
are free of gluonic corrections.
\vskip 0.7 cm
\noindent {\bf 4. Application to $K\ri 3\pi$ Decay}

    As stressed in Introduction, it is necessary to introduce a weak chiral
Lagrangian with higher derivatives in order to account for non-vanishing
quadratic coefficients and the discrepancy between theory and experiment
for the constant and linear terms in the Dalitz expansion of $K\ri 3\pi$
amplitude [Eq.(1.1)]. For $\di$ amplitudes, we find
a remarkable agreement between $1/N_c$ theory and experiment within $3\%$
for the constant and linear terms.
(The reader is referred to ref.[8] for more details.)
This means that very little room is left for chiral-loop corrections.
The predicted coefficient $c$ is just marginally in accord with data
within the experimental errors, while the other coefficient
$d$ is off by three standard deviations. Clearly more
accurate $K3\pi$ data are urgent to clarify this discrepancy.
In the $\dit$ sector, we see that the linear term is
generally in agreement with data, whereas the constant term is
four standard deviations off from experiment. Obviously, more
high-statistics
experiments are required to improve the determination of $\dit$
coefficients $a$ and $b$, and to extract the quadratic
terms $c$ and $d$ in order to test the chiral-Lagrangian approach.

   Since the quadratic slope paremeter in the $K_L\ri 3\pi^0$ Dalitz
plot was recently measured at Fermilab based on a sample of $5.1\times
10^6$ decays,\refmark{22} it is very interesting to compare
the $1/N_c$ prediction with experiment. The isospin structure of the
$K_L\ri 3\pi^0$ Dalitz amplitude is given by
$$A(K_L\ri 3\pi^0)=-3(a_1-2a_3)-3(c_1-2c_3)(Y^2+X^2/3),\eqno(4.1)$$
where the subscript 1 (3) refers to $\di~({3\over 2})$ transition. The
quadratic slope parameter $h$ for the decay is $2(c_1-2c_3)/(a_1-2a_3)$.
{}From Tables 1 and 2 of ref.[8] we find
$$h= -4.7\times 10^{-3},\eqno(4.2)$$
in accord with the result from the Fermilab E731 experiment
$$h_{{\rm expt}}=-(3.3\pm 1.1\pm 0.7)\times 10^{-3}.\eqno(4.3)$$
For comparsion, a somewhat large value of $-(1.2\pm 0.4)\times 10^{-2}$
for $h$ is predicted by ref.[2].

  Finally, we would also like to mention the decay $K_S\ri \pi^+
\pi^-\pi^0$, whose Dalitz amplitude is of the form $X(1+\alpha Y)$.
Explicitly,\refmark{8}
$$A(K_S\ri\pi^+\pi^-\pi^0)=-{2\over 3}b'_3X+{4\over 3}d'_3XY.\eqno(4.4)$$
We predict that\refmark{8}
$$Br(K_S\ri\pi^+\pi^-\pi^0)=3.9\times 10^{-7},~~~\alpha=4.4\times
10^{-2}.\eqno(4.5)$$
The experimental feasibility for measuring this decay mode is not
pessimistic.

\vskip 0.7 cm
\noindent {\bf 5. Applications to Radiative Kaon Decay}

\noindent {5.1~~The $K^+\ri\pi^+\gamma\gamma$ decay}

As first pointed out by EPR,\refmark{17} the loop amplitudes
of $K_{L,S}\ri\pi
\gamma\gamma$ and $K^+\ri\pi^+\gamma\gamma$ are finite. From the point
of view of large \nc chiral-Lagrangian approach, the mode $K^+\ri\pi^+
\gamma\gamma$ is more interesting since it also receives contributions
from the tree Lagrangians $\clnon$ and $\clf_S$ via pole diagrams
(except for the $\omega_3$ term which contributes via
 the direct-emission diagram).
  The total decay rate of $K^+\ri\pi^+\gamma\gamma$ was calculated
in ref.[17] to be
$$\Gamma(K^+\ri\pi^+\gamma\gamma)=\Gamma_{\rm loop}+\Gamma_{\rm tree}
+\Gamma_{\rm int}+\Gamma_{_{\rm WZW}},\eqno(5.1)$$
with
$$\eqalign{\Gamma_{\rm loop}= &\,2.80\times 10^{-23}{\rm GeV},~~
\Gamma_{\rm tree}=0.17\hat{c}^2\times 10^{-23}{\rm GeV}, \cr
\Gamma_{\rm int}= &\,0.87\hat{c}\times 10^{-23}{\rm GeV},~~
\Gamma_{_{\rm WZW}}=0.26\times 10^{-23}{\rm GeV}, \cr}\eqno(5.2)$$
and
$$\hat{c}=32\pi^2\,[\,4(L_9+L_{10})-{1\over 3}(\omega_1+2\omega_2
+2\omega_3)\,].\eqno(5.3)$$
Note that the combinations $\omega_1+2\omega_2+2\omega_3$ and $L_9+
L_{10}$ are separately scale independent. From
Eqs.(2.4) and (3.6) we obtain $\hat{c}=-4$ and the branching ratio
$$Br(K^+\ri\pi^+\gamma\gamma)=\,5.1\times 10^{-7}.\eqno(5.4)$$

   Since this decay is dominated by chiral-loop effects, its decay rate
is rather insensitive to the model of higher-derivative chiral Lagrangians.
For example, $\hat{c}$ is predicted to be zero in the so-called ``weak
deformation model",\refmark{23} but the corresponding branching ratio
$5.8\times 10
^{-7}$ is very close to that in the $1/N_c$ approach. In order to
discriminate these two models, experimentally it is important to measure the
two-photon spectrum around $z=m^2_{\gamma\gamma}/m^2_K=0.3$ where the
spectrum behaves quite differently for $\hat{c}=-4$ and 0 (see Fig.2 of
ref.[9]).

  The present best upper limit\refmark{24} $1.0\times 10^{-6}$ for
$K^+\ri\pi^+\gamma\gamma$ was obtained by assuming a $\pi^+$ energy
distribution given by phase space. If the theoretical spectrum for $\hat{c}
=-4$ is used, then the upper limit will be pulled
back to the level of $1.5\times 10^{-4}$.\refmark{24}

\noindent {5.2~~Direct $K\ri\pi\pi\gamma$ transitions}

The structure-dependent photon-emission decay $K\ri\pi\pi\gamma$ provides
an excellent probe on the $p^4$ weak chiral Lagrangian coupled to
external electromagnetic fields.
Under Lorentz and gauge invariance, the general expression for the
invariant direct emission (DE) amplitude of the
decay $K(k)\ri\pi(p_1)\pi(p_2)\gamma(q)$ reads
$$\eqalign{ & A_\de= \,\tilde{\beta}M+\tilde{\gamma}E, \cr
M\equiv &\,e\epsilon_{\mu\nu\rho\sigma}p_1^\mu p_2^\nu q^\rho \epsilon
^\sigma,~~~E\equiv\,e[\,(p_1\cdot\epsilon)(p_2\cdot q)-(p_2\cdot
\epsilon)(p_1\cdot q)\,],  \cr}\eqno(5.5)$$
where $\epsilon_\mu$ is the polarization vector of the photon.
The first term of $A_\de$ corresponds to magnetic transitions whereas
the second term is caused by electric transitions.
Evidently, the DE amplitude is of third power in momenta.
Taking into account the
experimental cutoff on the photon energy, we have the following
branching ratios
$$\eqalign{ \br(K^+\ri\pi^+\pi^0\gamma)_\de= &\,1.32\times 10^5
\GeV^6\,(|\tilde{\beta}|^2+|\tilde{\gamma}|^2), \cr \br(K_L\ri\pi^+
\pi^-\gamma)_\de= &\,1.33\times 10^6\GeV^6\,(|\tilde{\beta}|^2+
|\tilde{\gamma}|^2),  \cr \br(K_S\ri\pi^+\pi^-\gamma)_\de= &\,
2.28\times 10^3\GeV^6\,(|\tilde{\beta}|^2+|\tilde{\gamma}|^2).
\cr}\eqno(5.6)$$

   There are two contributions to direct photon emission of $K\ri
\pi\pi\gamma$: long-distance pole contributions and contact-term
ones (i.e., direct weak transitions). The long-distance
pole contribution is governed by the anomalous Wess-Zumino-Witten
interaction. Note that in the limit of $CP$ symmetry, $K_L\ri\pi^+\pi^-
\gamma$ proceeds only via the magnetic transition, whereas $K_S\ri\pi^+
\pi^-\gamma$ is caused by electric transition. At first sight, one may
tempt to conclude that the theoretical prediction for $K_L\ri\pi^+\pi^-
\gamma$ should be most reliable because it is entirely determined by
chiral anomalies. We will see later that it is not the case.

  Numerical values of the $1/N_c$ predictions are shown
in Table I of ref.[9]. It is
evident from Table I that the agreement between theory and experiment
for the direct emission  of $K^+\ri\pi^+\pi^0\gamma$
is striking, implying that very little room is left
for chiral-loop corrections. For $K_S\ri\pi^+\pi^-\gamma$, the branching
ratio of the structure-dependent component is predicted to be
$2\times 10^{-7}$, which
is beyond the present upper limit\refmark{25} $6\times 10^{-5}$.

  We cannot make a definite prediction for the direct emission
of $K_L\ri\pi^+\pi^-\gamma$ owing to a large theoretical uncertainty
in the estimate of the long-distance effect, as we shall discuss shortly.
In the absence of the $\eta'$ pole, it is easily seen that the pole
contribution to $K_L\ri\pi^+\pi^-\gamma$ vanishes due to the Gell-Mann-Okubo
mass relation $m^2_\eta={1\over 3}(4m^2_K-m^2_\pi)$. However,
the direct weak
contribution alone will yield a branching ratio of $2\times 10^{-4}$,
 which is too large by an order of magnitude when compared with experimental
branching fraction\refmark{26} of $(2.89\pm 0.28)\times 10^{-5}$.
This means that a large {\it destructive} interference
between pole and direct-transition amplitudes of $K_L\ri\pi^+\pi^-
\gamma$ is required in order to explain data. The $\eta'$ pole is thus
called for.

    The inclusion of the $\eta'$ intermediate state introduces two
complications: First,
the $SU(3)$ singlet $\eta_0$ is outside of the framework of $SU(3)\times
SU(3)$ ChPT; that is, the matrix element $\bra{\eta_0}\cl\ket{K_L}$ is
not related to the $K_L-\pi^0$ transition by $SU(3)$ symmetry. Second,
there will be an $\eta-\eta'$ mixing effect. In the $U(3)\times U(3)$
version of $\cl$, the above-mentioned two matrix elements are
connected via nonet symmetry, viz.
$$\bra{\eta_0}\cl\ket{K_L}=-2\sqrt{2\over 3}\rho\bra{\pi^0}\cl\ket{
K_L},\eqno(5.7)$$
where the parameter $\rho$ is introduced so that the deviation of
$\rho$ from unity implies the breakdown of nonet symmetry.
The pole contribution is quite sensitive to $SU(3)$ symmetry and nonet
symmetry breaking. For example, in the absence of symmetry breaking
effects the branching ratio is calculated to be $6.4\times 10^{-5}$,
which is two times large. Neglecting $SU(3)$ breaking and fitting to
the experimental centeral value, we find $\rho\approx 1.1$. This illustrates
that presently no definite prediction on the pole effects can be made
with certainty.
\foot{Efforts of relating the pole contributions of $K_L\ri\pi^+\pi^-
\gamma$ and $K_L\ri\gamma\gamma$ have been made before (see e.g., refs.[9,
20,27]). However, as emphasized by Shore and Veneziano,\refmark{28} the
naive PCAC analysis does not apply to the decay $\eta'\ri\gamma\gamma$.}

Finally, two remarks are in order. (i) Pole and contact-term
contributions are equally important for the direct radiative
transition of $K^+$ and $K_L$, whereas only the latter one
contributes to $K_S\ri\pi^+\pi^-\gamma$ in the limit of CP
symmetry. (ii) Unlike inner bremsstrahlung, the direct-emission
amplitudes of $K^\pm\ri\pi^\pm\pi^0\gamma$ and $K_L\ri\pi^+\pi^-\gamma$ are
no longer subject to the $\di$ rule and CP violation, respectively.
This explains why the branching ratio of $K^+$ and $K_L$ is larger
than that of $K_S$ by two orders of magnitude and why
structure-dependent effects can be seen in those two decay modes.
\vskip 0.7 cm
\noindent {\bf 6.~~Conclusion}

We have applied the large \nc chiral Lagrangian
approach to nonleptonic and radiative kaon decays. Decay rates and
spectra are
unambiguously predictable to the leading $1/N_c$
expansion and to the zeroth order in gluonic modifications.
Future high-statistics experiments with great
sensitivity will be able to test those predictions.
\vskip 2.0 cm
\centerline{ACKNOWLEDGMENTS}

   The author wishes to thank Profs. K. Goeke and W-Y. Pauchy Hwang
for inviting him to participate in this Symposium.
   This work was supported in part by the National Science Council
of the Republic of China.

\def\jo{\journal}
\def\pr{Phys. Rev.}
\def\prl{Phys. Rev. Lett.}
\def\pl{Phys. Lett.}
\def\np{Nucl. Phys.}
\def\zp{Z. Phys.}
\def\tr{{\rm Tr}}

\def\et{{\it et al.}}

\ref{J. Gasser and H. Leutwyler, {\sl Nucl. Phys.} {\bf B250} (1985) 465.}
\ref{J. Kambor, J.F. Donoghue, B.R. Holstein, J. Missimer, and D. Wyler
\jo\prl&68(92)1818.}
\ref{J. Balog, {\sl Phys. Lett.} {\bf B149} (1984) 197;
{\sl Nucl. Phys.} {\bf B258} (1985) 361;
K. Seo, Gifu preprint GWJC-1, 1984 (unpublished); A.A.
Andrianov\jo\pl&157B(85)425.}
\ref{R. MacKenzie, F. Wilczek, and A. Zee\jo\prl&53(84)2203; I.J.R.
Aitchison and C.M. Fraser\jo\pl&146B(84)63; G. Bhattacharya and S.
Rajeev, in {\it Proceedings of the Sixth Annual
Montreal-Rochester-Syracuse-Toronto Meeting, Syracuse, 1984}
(Syracuse University, Syracuse, New York, 1984);
L.-H. Chan\jo\prl&55(85)21; P. Simi\'c\jo ibid.&55(85)40; A. Zaks\jo
\np&B260(85)241; C. Fraser\jo Z. Phys.&C28(85)101.}
\ref{D. Espriu, R. de Rafael, and J. Taron\jo\np&B345(90)22.}
\ref{D. Ebert and H. Reinhardt\jo\np&B271(86)188; for a recent similar
derivation of the low-energy action of the Nambu-Jona-Lasinio model, see
J. Bijnens, C. Bruno, and E. de Rafael, CERN-TH-6521/92.}
\ref{H.Y. Cheng\jo\pr&D42(90)3850.}
\ref{H.Y. Cheng\jo\pl&B238(90)399.}
\ref{H.Y. Cheng\jo\pr&D42(90)72.}
\ref{J.A. Cronin\jo\pr&101(67)1483.}
\ref{Particle Data Group\jo\pr&D45(92)S1.}
\ref{Yu. M. Antipov {\it et al.}\jo\zp&C26(85)495; {\sl Phys. Lett.}
{\bf 121B} (1983) 445.}
\ref{G. Ecker, J. Gasser, A. Pich, and E. de Rafael\jo\np&B321(89)311.}
\ref{E. Golowich\jo\pr&D35(87)2764; {\sl ibid.} {\bf D36} (1987) 3516.}
\ref{J. Kambor, J. Missimer, and D. Wyler\jo\np&B346(90)17.}
\ref{W.A. Bardeen, A.J. Buras, J.-M. G\'erard\jo\pl&192B(87)138;
{\sl Nucl. Phys.} {\bf B293} (1987) 787;
S. Fajfer and J.-M. G\'erard\jo\zp&C42(89)425.}
\ref{G. Ecker, A. Pich and E. de Rafael,  {\sl Nucl. Phys.} {\bf
B291} (1987) 692; {\sl Phys. Lett.} {\bf B189} (1987) 363;
{\sl Nucl. Phys.} {\bf B303} (1988) 665.}
\ref{C. Alliegro {\it et al.}\jo\prl&68(92)278.}
\ref{C. Bruno and J. Prades, CPT-92/P.2795 (1992).}
\ref{G. Ecker, H. Neufeld, and A. Pich\jo\pl&B278(92)337.}
\ref{J. Bijnens, G. Ecker, and A. Pich\jo\pl&B286(92)341.}
\ref{S.V. Somalwar {\it et al.}\jo\prl&68(92)2580.}
\ref{G. Ecker, A. Pich, and E. de Rafael\jo\pl&B237(90)481.}
\ref{M.S. Atiya {\it et al.}\jo\prl&65(90)1188.}
\ref{Experiments for $K_S\ri\pi^+\pi^-\gamma$: H. Taureg \et\jo\pl&
65B(76)92; G. Burgun \et\jo\pl&46B(73)481.}
\ref{Experiment for $K_L\ri\pi^+\pi^-\gamma$: A.S. Carroll \et\jo\prl
&44(80)529.}
\ref{H.Y. Cheng\jo\pl&B245(90)122.}
\ref{G.M. Shore and G. Veneziano\jo\np&B381(92)3.}
\endpage
\refout
\end